\documentclass[aps,prd,preprint,superscriptaddress,showpacs]{revtex4}
\usepackage{amsmath}
\usepackage{amssymb}
\usepackage{epsfig}
\usepackage{ulem}
\usepackage{graphics}
\usepackage{indentfirst}
\usepackage{color}
\usepackage{multirow}

\newcommand{\tb}{\ensuremath{\bar{T}}}
\newcommand{\kb}{\ensuremath{\bar{\chi}}}
\newcommand{\pb}{\ensuremath{\bar{\phi}}}
\newcommand{\tq}{\ensuremath{\tilde{q}}}
\newcommand{\lml}{\ensuremath{\lambda_1}}
\newcommand{\lmll}{\ensuremath{\lambda_2}}

\newcommand{\wbl}{\ensuremath{\bar{w}_1}}
\newcommand{\wbll}{\ensuremath{\bar{w}_2}}
\newcommand{\wnp}{\ensuremath{W_{np}}}
\newcounter{RomanNumber}

\begin{document}

\preprint{ACT-8-14, MIFPA-14-22}

\title{Aligned Natural Inflation and Moduli Stabilization from Anomalous $U(1)$ Gauge Symmetries}

\author{Tianjun Li \vspace*{0.1cm}}
\affiliation{{\footnotesize State Key Laboratory of Theoretical Physics
and Kavli Institute for Theoretical Physics China (KITPC),
      Institute of Theoretical Physics, Chinese Academy of Sciences,
Beijing 100190, P. R. China}}

\affiliation{{\footnotesize School of Physical Electronics,
University of Electronic Science and Technology of China,
Chengdu 610054, P. R. China}}

\author{Zhijin Li}

\affiliation{{\footnotesize George P. and Cynthia W. Mitchell Institute for
Fundamental Physics and Astronomy,
Texas A\&M University, College Station, TX 77843, USA}}

\author{Dimitri V. Nanopoulos}

\affiliation{{\footnotesize George P. and Cynthia W. Mitchell Institute for
Fundamental Physics and Astronomy,
Texas A\&M University, College Station, TX 77843, USA}}

\affiliation{{\footnotesize Astroparticle Physics Group, Houston Advanced
Research Center (HARC), Mitchell Campus, Woodlands, TX 77381, USA}}

\affiliation{{\footnotesize Academy of Athens, Division of Natural Sciences,
28 Panepistimiou Avenue, Athens 10679, Greece} \vspace*{0.7cm}}

\begin{abstract}

To obtain natural inflation with large tensor-to-scalar ratio in string framework, we need
a special moduli stabilization mechanism which can separate the masses of real and imaginary
components of K\"ahler moduli at different scales, and achieve a trans-Planckian axion decay constant
from sub-Planckian axion decay constants. In this work, we stabilize the matter fields by F-terms
and the real components of K\"ahler moduli by D-terms of two anomalous $U(1)_X\times U(1)_A$
symmetries strongly at high scales, while the corresponding axions remain light due to their independence on
the Fayet-Iliopoulos (FI) term in moduli stabilization. The racetrack-type
axion superpotential is obtained from gaugino condensations of the hidden gauge symmetries
$SU(n)\times SU(m)$ with massive matter fields in the bi-fundamental respresentations.
The axion alignment via Kim-Nilles-Pelroso (KNP) mechanism corresponds to an approximate $S_2$ exchange
symmetry of two K\"ahler moduli in our model, and a slightly $S_2$ symmetry breaking leads
to the natural inflation with super-Planckian decay constant.

\end{abstract}

\pacs{04.65.+e, 04.50.Kd, 12.60.Jv, 98.80.Cq}

\maketitle

\section{Introduction}

Natural inflation was proposed to explain the unnatural flatness of inflationary potential
which is introduced {\itshape ad hoc} at tree level and remains flat under radiative
corrections \cite{Freese:1990rb}. The flatness of inflationary potential is protected
by continuous shift symmetry of an axionic field $\phi$. But this symmetry is
spontaneously broken to a discrete shift symmetry at inflation scale, and the following
inflationary potential is generated
\begin{equation}
V(\phi)=\Lambda^4(1\pm{\rm cos}(\frac{\phi}{f})), \label{V}
\end{equation}
which is invariant under the discrete shift symmetry $\phi\rightarrow\phi+2\pi f$
with $f$ as an axion decay constant.
Recent observation of the B-mode polarization by the BICEP2 Collaboration
suggests a large tensor-to-scalar ratio $r=0.16^{+0.06}_{-0.05}$ excluding
the dust effects~\cite{Ade:2014xna}, which can be obtained from natural inflation with
trans-Planckian axion decay constant $f\sim O(10) ~{\rm M_{Pl}}$,
 where ${\rm M_{Pl}}=2.4\times10^{18} {\rm GeV}$ is the reduced Planck mass~\cite{Freese:2014nla}.
While the value of $r$ could be much smaller \cite{Mortonson:2014bja, Flauger:2014qra} if the dust polarization effect plays more important role than estimated in Ref. \cite{Ade:2014xna}.
Besides, natural inflation with large $r$ also agrees with the
Planck observations \cite{Ade:2013uln}, in which a lower bound of trans-Planckian axion decay constant
is needed $f\geqslant 5 {\rm M_{Pl}}$.

It is an attractive destination to realize natural inflation in string theory or its efffective
no-scale supergravity (SUGRA) theory. Axions arise from anti-symmetric tensor fields in string theory
through compactification of the extra space dimensions, and may play important roles in
cosmology and particle physics~\cite{Svrcek:2006yi}. A more fundamental reason for stringy inflation is that,
according to the Lyth bound obeyed by general single field slow-rolling process \cite{Lyth:1996im},
large tensor-to-scalar ratio requires trans-Planckian excursion during inflation. Therefore,
 corrections from the Planck-suppressed operators are non-ignorable, and
a reliable inflation theory has to be constructed based on the ultra violet (UV) complete theory
such as string theory.

Axion (as imaginary component of K\"ahler modulus $T$) inflation in string theory needs a moduli stabilization
mechanism, which can separate the masses of real and imaginary components of K\"ahler modulus $T$ at
different scales. Specifically, the real component of modulus should be frozen  during inflation,
so it obtains a large mass from modulus stabilization: $M_{\rm Re{(T)}}> H$, where $H$ is the Hubble
constant during inflation, at the order $10^{14}$ GeV from the BICEP2 results, while
the ``effective mass'' of axion $M_{\rm Im(T)}=\Lambda^2/f$ is of order $10^{13}$ GeV.
In the well-known KKLT mechanism \cite{Kachru:2003aw}, once the real component of modulus
is stabilized, the imaginary component obtains a large mass comparable to
the real component which destroys axion inflation \cite{Kallosh:2007ig}.

Moduli stabilization, which is consistent with axion inflation, has been proposed recently
in Refs.~\cite{Li:2014owa, Li:2014xna}. In these works, an anomalous $U(1)_X$ gauge symmetry
has been introduced to split the masses of real and imaginary components of K\"ahler moduli.
The anomalous $U(1)_X$ gauge symmetry introduces the moduli-dependent Fayet-Iliopoulos (FI) term as a result of
the non-trivial moduli transformation under $U(1)_X$. The D-term scalar potential of $U(1)_X$
depends only on the real component of moduli, and actually is close to string scale by taking suitable
gauge charges. Consequently, the D-term flatness provides a strong stabilization
on real component of moduli once the $U(1)_X$ charged matter fields are stabilized by F-terms.
The stabilization of matter fields can be directly done by tree-lever superpoential.
While axion potential for natural inflation is obtained from non-perturbative effects, so it is
generally much weaker than the perturbative terms \cite{Li:2014xna}. In the model, besides moduli
stabilization consistent with axion inflation, $U(1)_X$ symmetry also provides an elegant solution
to the problem of super-Planckian axion decay constant. A general review on the anomalous $U(1)$ gauge symmetry in SUGRA and its applications on cosmology is provided in Ref.~\cite{Binetruy:2004hh}.

However, the super-Planckian axion decay constant in natural inflation required by the BICEP2 results is
problematic in string theory. String theory predicts the axion decay constants cannot surpass
the string scale \cite{Svrcek:2006yi, Choi:1985je, Banks:2003sx}, while as a controllable theory,
the scale of weakly coupled string theory should be lower than the Planck scale. The
Kim-Nilles-Pelroso (KNP) mechanism was proposed to obtain the effective super-Planckian axion decay
constant~\cite{Kim:2004rp}. In this proposal, two axions with sub-Planckian decay constants are
aligned to have a flat direction along which the effective decay constant can be large enough
for natural inflation. Another solution to this problem was provided in Ref.~\cite{Li:2014xna} based on
an anomalous $U(1)$ gauge symmetry. In this work the axion decay constant is directly determined by
the flatness of anomalous $U(1)$ D-term, and can be super-Planckian by taking a reasonably
large condensation gauge group.

Recently,  a lot of works on the KNP mechanism have been done after the BICEP2
results~\cite{Choi:2014rja, Higaki:2014pja, Kappl:2014lra, Long:2014dta, Gao:2014uha}. Specifically,
in Ref.~\cite{Long:2014dta} a stringy geometrical realization of aligned axions was proposed based on
the assumption that the moduli are well stabilized. While a complete realization of the KNP mechanism
based on string framework, which is consistent with moduli stabilization, is still absent.

In this paper, we will realize both the KNP mechanism and moduli stabilization in string inspired
no-scale supergravity with anomalous gauge symmetries $U(1)_X\times U(1)_A$. Similar to our previous
study~\cite{Li:2014owa, Li:2014xna}, we stabilize the matter fields by F-term potential and the moduli
 by the D-terms of anomalous $U(1)$ symmetries. Two axions are imaginary components of two K\"ahelr
moduli, which transform non-trivially under two anomalous gauge symmetries $U(1)_X\times U(1)_A$.
The axion superpotential is of racetrack-type, and is from gaugino condensations of two gauge groups
$SU(n)\times SU(m)$ with massive quark representations. The alignment of axions as in the KNP proposal,
actually corresponds to an approximate $S_2$ exchange symmetry between two K\"ahler moduli.
This $S_2$ symmetry is explicitly broken slightly so that one linear combination of
the axions gives us the natural inflation with
 super-Planckian effective decay constant.

This paper is organized as follows. In Section 2 we present string inspired no-scale SUGRA with
anomalous gauge symmetries $U(1)_X\times U(1)_A$. In Section 3 we provide stabilizations of matter fields
and K\"ahler moduli based on F-term and D-term potentials, respectively. In Section 4 we show that
the aligned natural inflation can be realized by superpotential obtained from gaugino condensations
of hidden gauge groups. Conclusions are given in Section 5.

\section{Aligned axions in no-scale SUGRA}

We start from the no-scale type SUGRA with K\"ahler potential
\begin{equation}
K=-{\rm ln}(T_1+\tb_1)-{\rm ln}(T_2+\tb_2)+\phi_i\pb_i+\chi_i\kb_i+\psi_i\bar{\psi}_i+X_i\bar{X}_i+Y_i\bar{Y}_i~,
\end{equation}
where $i=1, 2$, and $T_1$ and $T_2$ are K\"ahler moduli.
No-scale SUGRA \cite{Cremmer:1983bf} was realized
naturally in the compactifications of weakly coupled heterotic string
theory \cite{Witten:1985xb} or M-theory on $S^1/Z_2$ \cite{Li:1997sk}.
For the third K\"ahler modulus $T_3$, we assume that it is neutral under anomalous gauge symmetries
$U(1)_X\times U(1)_A$ and then is ignored in this work.
Two axions $\theta_i$ are the imaginary parts of the K\"ahler moduli, {\it i.e.},
 $\theta_i\equiv {\rm Im}(T_i)$, $i=1, 2$.

As we know, $n$ stacks of D7-branes, which wrap a 4-cycles of the Calabi-Yau space, gives $U(n)$ gauge group.
As in the KKLT scenario, the condensation gauge group is $SU(n)$, and typically there is another
anomalous $U(1)$ gauge symmetry. For two K\"ahler moduli, there can be two copies of such gauge sectors, saying
$SU(n)\times SU(m)\times U(1)_X\times U(1)_A$. In particular, two anomalous $U(1)$ gauge symmetries can
play special role in moduli stabilization and inflation.

We assume that the vector-like massive quarks are in the fundamental representations of Yang-Mills gauge groups
$SU(n)\times SU(m)$. After integrating out the heavy chiral superfields, the gaugino condensations
of gauge groups $SU(n)\times SU(m)$ generate the following effective superpotential
\begin{equation}
W=A\phi_1^{\frac{1}{m}}e^{-(aT_1+bT_2)}+B\phi_2^{\frac{1}{n}}e^{-(aT_1+bT_2+cT_2)}, \label{nsp}
\end{equation}
where $\phi_1$ and $\phi_2$ are matter fields charged on both
$SU(n)$ and $SU(m)$ \cite{Taylor:1982bp, Lust:1990zi, deCarlos:1991gq}. Here, we have assumed that the gauge kinetic functions of $SU(n)$ and $SU(m)$, which relate to the gauge anomaly cancellations of $SU(n)^2\times U(1)_a$ and $SU(m)^2\times U(1)_a$, are
\begin{align}
&f_{SU(n)}\propto aT_1+bT_2+cT_2~, \\
&f_{SU(m)}\propto aT_1+bT_2~.
\end{align}
 The superpotential in Eq.~(\ref{nsp}) and moduli dependent D-term are employed to
stabilize the matter fields and real parts of the moduli, respectively. Once the matter
fields $\phi_i$ obtain vacuum expectation values (VEVs), the real part of the K\"ahler moduli
are fixed by the D-term flatnesses. The
effects of non-perturbative terms on moduli stabilization and inflation based on anomalous $U(1)$ have been studied
in Refs.~\cite{Dudas:2005vv, Villadoro:2005yq, Achucarro:2006zf, Lalak:2005hr, Brax:2007fe, Lalak:2007qd}. In these works,
normally the D-terms are non-cancellable so that they can uplift the AdS vacua to dS vacua from moduli stabilization.
The non-cancellability of anomalous $U(1)$ D-term arises from the massless fundamental representation of condensation
gauge group. While here the cancellable anomalous $U(1)$ D-terms are preferred, the condensation gauge groups
are equipped with massive fundamental representations. Recently, under the stimulation from the BICEP2 results,
the potential roles of gaugino condensation on inflation have been studied in \cite{Dine:2014hwa, Yonekura:2014oja}.

In our strategy, fields $\phi_i$ are stabilized by their superpotential terms, which break anomalous $U(1)_X$
(also the continous shift symmetry of K\"ahler modulus) spontaneously. This procedure was first proposed
in Ref.~\cite{Li:2014xna} for string-inspired no-scale SUGRA, and later it was applied to the case of
the minimal SUGRA \cite{Mazumdar:2014bna}.

\subsection{$U(1)_X\times U(1)_A$ Gauge Invariance}

The $U(1)_a$ charges of the K\"ahler moduli and matter fields are provided in Table \ref{ta}, so
the overall superpotential, which is invariant under gauge transformations of $U(1)_a$ with $a\in \{A, X\}$,
is given by
\begin{equation}
\begin{split}
W=w_\star+A\phi_1^{\frac{1}{m}}e^{-(aT_1+bT_2)}+B\phi_2^{\frac{1}{n}}e^{-(aT_1+bT_2+cT_2)} ~~~\\
+X_i(\phi_i\chi_i-\lml)+Y_i(\psi_i\chi_i-\lmll)+c_0\psi_i\chi_i, \label{sp}
\end{split}
\end{equation}
in which the constant term $w_\star$ is arising after integrating out all complex-structure moduli.
Similar to Ref.~\cite{Li:2014xna}, we may realize the above superpotential or its equivalent.

\begin{table}
\begin{tabular}{l|llllllllll}
\hline
\hline
~& ~~~$T_1$ & $T_2$~~ & $X_i$~ & $Y_i$~ & $\phi_1$~ & $\phi_2$~ & ~$\chi_1$~ & $\chi_2$~ & ~$\psi_1$~ & $\psi_2$~~  \\
\hline
~~$U(1)_X$~~ & ~~~$\delta_X^1$~ & $\delta_X^2$~~ & ~0~ & 0~ & $q$~ & $q$~ & $-q$~ & $-q$~ & ~$q$ &  $q$~ \\
~~$U(1)_A$~~ & ~~~$\delta_A^1$~ & $\delta_A^2$~~ & ~0~ & 0~ & $\tq$~ & 2$\tq$~ &  $-\tq$~ &  $-2\tq$~ &  ~$\tq$ &  $2\tq$~  \\
\hline
\hline
\end{tabular}
\caption{$U(1)_X\times U(1)_A$ charges of the K\"ahler moduli and matter fields.} \label{ta}
\end{table}
The matter fields $z_n\in\{\phi_i, \chi_i, \psi_i\}$ transform linearly under $U(1)_a$, while K\"ahler moduli shift under the $U(1)_a$ gauge transformations
\begin{align}
\begin{split}
&T_i\rightarrow T_i+i\delta^i_a \epsilon, \\
&z_n\rightarrow z_n e^{i\epsilon q^a_{z_n}}.
\end{split}
\end{align}

The $U(1)_a$ gauge invariance of matter couplings in (\ref{sp}) is clearly based on the charges provided in Table \ref{ta}.
While for the non-perturbative terms, gauge invariance requires the following condition
\begin{displaymath}
\left( \begin{array}{cc}
a & b \\
a& b+c
\end{array} \right)
\left( \begin{array}{c}
\delta^1_\alpha  \\
\delta^2_\alpha
\end{array}
\right)=
\left( \begin{array}{c}
\frac{q^1_\alpha}{m} \\
\frac{q^2_\alpha}{n}
\end{array}
\right),
\end{displaymath}
where $\alpha\in \{A, X\}$ and $q^i_\alpha$ means the $U(1)_\alpha$ charge of $\phi_i$.
Without $c$ the equation is degenerate and the moduli charges cannot be uniquely determined based on charges of the matter fields.

In the superpotential (\ref{sp}), the parameter $c$ is very small.
There is an exact $S_2$ exchange symmetry
\begin{equation}
aT_1 \leftrightarrow bT_2,
\end{equation}
in K\"ahler potential $K$ and also in superpotential $W$ without term $cT_2$.
The exact $S_2$ symmetry corresponds to exact aligned axions, then the potential is independent with $bT_1-aT_2$, which becomes an exact flat direction in the scalar potential. The $S_2$ symmetry is broken explicitly by $cT_2$ while reserves approximately as long as $c\ll b$. This approximate symmetry is crucial that it provides sufficient flat potential for inflation.

In this work, we will take $m=n+1$ for simplicity, besides, the approximate $S_2$ symmetry is useful to simplify calculations.

Given the charges of $\phi_i$ in Table \ref{ta}, the $U(1)_a$ charges of moduli $T_i$ are uniquely fixed
\begin{align}
\begin{split}
&\delta_X^1=\frac{q}{a(n+1)}(1-\frac{b}{cn}), ~~~~~~~~~~~\delta_X^2=\frac{q}{cn(n+1)}, \\
&\delta_Y^1=\frac{\tq}{a(n+1)}(1-\frac{b(n+2)}{cn}), ~~~~\delta_Y^2=\frac{\tq(n+2)}{cn(n+1)}.  \label{del}
\end{split}
\end{align}
In this work the degree of gauge group $n$ is taken as $O(10)$, while the ratio $b/c$ is close to $O(10^2)$, so we have
$1-\frac{b}{cn}<0$, {\it i.e.}, the $U(1)_a$ charges of modulus $T_1$ are negative in unit $q$ or $\tq$.
Negative charges of $T_1$ are directly determined by the smallness of ratio $c/b$, i.e., the approximate $S_2$ symmetry between two K\"ahler moduli. This property is greatly appreciated in quantum anomaly cancellation, as will be shown later.

\subsection{Quantum Anomaly Cancellation of $U(1)_X\times U(1)_A$}

For a consistent gauge theory, all quantum anomalies associated with $U(1)_a$ should be cancelled. For gauge sectors $U(1)_X\times U(1)_A$, the gauge anomalies contain the cubic $U(1)^3_a$ anomaly, the gravitational $U(1)_a$ anomaly and the mixed $U(1)^2_a\times U(1)_b$ anomaly. The overall fermionic contributions on these gauge anomalies are non-zero, to keep theory free of gauge anomaly, the Green-Schwarz contributions on gauge anomaly \cite{Green:1984sg} are necessary.

For gravitational $U(1)_a$ gauge anomaly,  the fermionic contributions are
\begin{align}
\begin{split}
&{\rm Tr}~ q_X=\sum_z q_z=2q, \\
&{\rm Tr}~ q_A=\sum_z \tq_z=3\tq.
\end{split}
\end{align}
The gravitational anomalies are cancelled by higher derivative terms $R^2$. The mixed  anomalies such as $U(1)^2_a\times U(1)_b$
and $SU(n)\times U(1)_a$ can be cancelled as well.

For cubic $U(1)^3_a$ anomalies, the fermionic contributions are
\begin{align}
&{\rm Tr}~ q^3_X=\sum_z q_z^3=2q^3, \\
&{\rm Tr}~ q^3_A=\sum_z \tq_z^3=9\tq^3.
\end{align}
Gauge kinetic functions of $U(1)_a$ are
\begin{align}
f_{U(1)_a}=k_a^1T_1+k_a^2T_2, \label{gkf}
\end{align}
in which $k_a^1$ are positive parameters. Consequently, the gauge kinetic terms are
\begin{equation}
\int d^2\theta ~k_a^iT_iW_a^2~, \label{kg}
\end{equation}
where $W_a$ is the $U(1)_a$ gauge field strength. There are two parts in the gauge kinetic term: $Re(f)F^2$ and $Im(f)F\tilde{F}$. The first part is $U(1)_a$ invariant, while the second part shifts under $U(1)_a$ due to non-trivial $U(1)_a$ gauge transformations of K\"ahler moduli $T_i$, actually the $U(1)_a$ variation of gauge kinetic term cancels the cubic gauge anomaly from fermionic contributions. Vanishing of cubic $U(1)_a^3$ anomaly
requires
\begin{equation}
k_a^i\delta^i_a=-\frac{1}{48\pi^2}{\rm Tr}~ q^3_a,
\end{equation}
where an extra coefficient $1/3$ is introduced as a symmetry factor of $U(1)^3_a$ anomaly graphs.
Specifically, we have
\begin{align}
k_X^i\delta^i_X=-\frac{1}{24\pi^2} q^3, \\
k_A^i\delta^i_A=-\frac{3}{16\pi^2}\tq^3.
\end{align}
Here all the coefficients $k_a^i$ are positive. The above conditions cannot be fulfilled unless for each $U(1)_a$, at least one of the K\"ahler moduli has negative charge (with unit of $q$ or $\tq$). Fortunately, as shown in (\ref{del}),  we do have one K\"ahler modulus $T_1$ whose $U(1)_a$ charges are negative resulting from the approximate $S_2$ symmetry. Then for any values of $q/\tq$, it is easy to adjust the parameters $k_a^i$ so that the cubic $U(1)^3_a$ anomalies vanish.

\section{The K\"ahler Moduli and Matter Field Stabilization}

The matter fields are stabilized by F-term potential. Once the matter fields obtain VEVs, the real components of the moduli are fixed by vanishing of $U(1)_a$ D-terms. Normally field stabilization happens at scale much higher than inflation scale. The
couplings between matter fields and inflation potential can only slightly modify the VEVs of matter fields, and more detailed analysis is presented in Ref.~\cite{Li:2014owa, Li:2014xna}. Here, we just ignore the scalar potential relating to inflation at this stage.

The F-term scalar potential is determined by the K\"ahler potential $K$ and superpotential $W$
\begin{equation}
V_F=e^K(K^{i\bar{j}}D_iW D_{\bar{j}}\bar{W}-3W\bar{W}), \label{vks}
\end{equation}
in which $K^{i\bar{j}}$ is the inverse of the K\"ahler metric $K_{i\bar{j}}=\partial_i\partial_{\bar{j}}K$ and $D_iW=W_i+K_iW$. The D-term scalar potential is given by
\begin{equation}
V_D=\frac{1}{2}D_aD^a,
\end{equation}
in which the gauge indices $a$ are raised by the form $[(Ref)^{-1}]^{ab}$, and $f$ is the gauge kinetic function.
The $D_a$ components are
\begin{equation}
D_a=iK_iX^i_a+i\frac{W_i}{W}X^i_a,
\end{equation}
where $X^i_a$ are the components of Killing vectors $X_a=X^i_a(\phi)\partial/\partial\phi^i$ which generate the isometries of the K\"ahler manifold that are gauged to form $U(1)_a$.
If the superpotential $W$ is gauge invariant instead of gauge covariant, the $D_a$ components reduce to
\begin{equation}
D_a=iK_iX^i_a.
\end{equation}
For the $U(1)_a$ charged matter fields $z_n$, they transform linearly under $U(1)_a$, and the Killing vectors linearly depend on the matter field $z_n$
\begin{equation}
X_a^{z_n}=iq^a_{z_n}z_n.
\end{equation}
For the K\"ahler moduli $T_i$, they shift under $U(1)_a$ gauge transformations, and the Killing vectors are
\begin{equation}
X_a^{T_i}=i\delta_a^i,
\end{equation}
which are purely imaginary constants.

\subsection{Matter Field Stabilization}

Considering the renormalizable matter couplings in (\ref{sp}), it is clear that the neutral matter fields $X_i$ and $Y_i$ have global minimum at the origin, while the charged matter fields $\phi_i$, $\chi_i$ and $\psi_i$ obtain non-vanishing VEVs. During inflation, these matter fields will evolve to the global minimum very fast in consequence of the F-term exponential factor $e^{z_n\bar{z}_n}$ and the large masses obtained from the matter couplings in (\ref{sp}).

Part of the F-term potential is
\begin{equation}
\begin{split}
V_{\rm m}=e^K(|\phi_i\chi_i&-\lml|^2+|\psi_i\chi_i-\lmll|^2+c^2_0(|\psi_i|^2+|\chi_i|^2) \\
         &+2c_0(\psi_i\chi_i\bar{W}+\bar{\psi}_i\kb_iW)+\cdots),
\end{split}
\end{equation}
where we have ignored the terms proportional to $X_i, Y_i$ or containing $\phi_i, \chi_i, \psi_i$ while several orders smaller.
The small term $2c_0(\psi_i\chi_i\bar{W}+\bar{\psi}_i\kb_iW)$, although ignorable for field stabilization, has considerable contribution to inflation potential.

As shown in \cite{Li:2014xna}, for $\lmll\gg c^2_0$, above potential admits a global minimum at
\begin{align}
\begin{split}
&|\psi_i|=|\chi_i|\simeq\sqrt{\lmll}, \\
&|\phi_i|\equiv r_i\simeq\frac{\lml}{\sqrt{\lmll}}.\label{vev}
\end{split}
\end{align}
The $U(1)_a$ gauge symmetries are broken spontaneously by non-zero VEVs.
Through matter field stabilization, potential $V_{\rm m}$ obtains VEVs as well and uplifts the vacuum energy
\begin{equation}
\langle V_{\rm m}\rangle=e^{\langle K\rangle}c^2_0(\langle\psi_i\rangle^2+\langle\chi_i\rangle^2)\simeq4c^2_0\lmll e^{\langle K\rangle}.
\end{equation}
Normally the non-perturbative superpotential associated with no-scale K\"ahler potential leads to the
AdS vacua. To obtain the Minkowski or dS vacua, an uplifting mechanism is needed.
Here, the positive vacuum energy obtained from matter field stabilization provides a natural solution to this problem.

Up to now we have ignored the effects of the
lower order terms on matter field stabilization. In \cite{Li:2014xna} these effects have been studied, it was shown that the small couplings can only lead to tiny corrections on these VEVs, and reduce the vacuum energy slightly, which are totally ignorable in a general estimation.

\subsection{Moduli Stabilization from $U(1)_X\times U(1)_A$ D-terms}
According to the $U(1)_a$ charges provided in Table \ref{ta}, the D-term potentials associated with $U(1)_a$ are
\begin{align}
V_{Da}=\frac{1}{2f_{U(1)_a}} (-\frac{\delta_a^1}{T_1+\tb_1}-\frac{\delta_a^2}{T_2+\tb_2}+q^a_{z_n}z_n\bar{z}_n)^2,
\end{align}
where $z_n\in \{\phi_i, \chi_i, \psi_i\}$ and $q^a_{z_n}$ is $U(1)_a$ charge of field $z_n$. Gauge kinetic function $f_{U(1)_a}$ is provided in (\ref{gkf}). The matter fields $z_n$ obtain VEVs through F-term stabilization, then the $U(1)_a$ D-terms become
\begin{equation}
\begin{split}
&V_{DX}=\frac{1}{2f_{U(1)_X}} (-\frac{\delta_X^1}{T_1+\tb_1}-\frac{\delta_X^2}{T_2+\tb_2}+qr_1^2+qr_2^2)^2,   \\
&V_{DA}=\frac{1}{2f_{U(1)_A}} (-\frac{\delta_A^1}{T_1+\tb_1}-\frac{\delta_A^2}{T_2+\tb_2}+\tq r_1^2+2\tq r_2^2)^2,
\end{split}
\end{equation}
which are vanished at vacuum. Together with the charges of moduli in (\ref{del}),
the real components of K\"ahler moduli $T_i\equiv T_{Ri}+i\theta_i$
can be uniquely determined at vacuum
\begin{equation}
\begin{split}
&\frac{1}{2a\langle T_{R1}\rangle}=(n+1)r^2_1+nr^2_2,  \\
&\frac{1}{2b\langle T_{R2}\rangle}=(n+1)r^2_1+nr^2_2+\frac{cn}{b}r^2_2. \label{ms}
\end{split}
\end{equation}
Even though the $U(1)_a$ gauge symmetries are broken after field stabilization, the approximate discrete symmetry $S_2$ is sustained. The symmetry breaking factor is very small, $c\ll b$,
therefore the VEVs of $T_i$ satisfy $a\langle T_{R1}\rangle\simeq b\langle T_{R2}\rangle\equiv r/2$ which is guaranteed by the approximate $S_2$ symmetry.

The imaginary components $\theta_i$ remain free in the perturbative potential, so actually they only appear in the potential through non-perturbative effects.

\section{Inflation Potential}
Field stabilization is happened at scale much higher than the inflation scale. After stabilization, we get the following effective superpotential
\begin{equation}
W=w_0+A\phi_1^{\frac{1}{m}}e^{-(aT_1+bT_2)}+B\phi_2^{\frac{1}{n}}e^{-(aT_1+bT_2+cT_2)},
\end{equation}
where $w_0=w_\star+2c_0\lmll$,
and there is a positive cosmology constant term $4c_0^2\lmll e^{\langle K\rangle}$, which is necessary to uplift the AdS vacua to Minkowski or dS vacua.

Denoting $W_{np}=A\phi_1^{\frac{1}{m}}e^{-(aT_1+bT_2)}+B\phi_2^{\frac{1}{n}}e^{-(aT_1+bT_2+cT_2)}\equiv w_1+w_2$, we have $W_{T_1}=-aW_{np}$, $W_{T_2}=-bw_1-(b+c)w_2\simeq-b\wnp$. The F-term scalar potential includes
\begin{equation}
\begin{split}
D_1WD_{\bar{1}}\bar{W}K^{1\bar{1}}=&a^2(T_1+\tb_1)^2\wnp\bar{W}_{np}    \\
&+a(T_1+\tb_1)(w_0(\wnp+\bar{W}_{np})+2\wnp\bar{W}_{np})+W\bar{W},
\end{split}
\end{equation}
and
\begin{equation}
\begin{split}
D_2WD_{\bar{2}}\bar{W}K^{2\bar{2}}=&b^2(T_2+\tb_2)^2\wnp\bar{W}_{np}    \\
&+b(T_2+\tb_2)(w_0(\wnp+\bar{W}_{np})+2\wnp\bar{W}_{np})+W\bar{W}.
\end{split}
\end{equation}
Therefore, parts of the moduli contributions on F-term potential are
\begin{equation}
V_T\propto2r^2\wnp\bar{W}_{np}+2r(w_0(\wnp+\bar{W}_{np})+2\wnp\bar{W}_{np})+2W\bar{W}, \label{VT}
\end{equation}
in which we have used the approximate $S_2$ symmetry $\langle aT_{R1}\rangle\simeq\langle bT_{R2}\rangle\equiv r/2$.

For the matter fields $\phi_i$, we have
\begin{equation}
\begin{split}
&W_{\phi_1}=\frac{w_1}{m\phi_1},  \\
&W_{\phi_2}=\frac{w_2}{n\phi_2},
\end{split}
\end{equation}
where the terms proportional to $X_i$ are ignored. The F-term potential contains
\begin{equation}
\begin{split}
V_{\phi_1}&=W_{\phi_1}{\bar{W}}_{\pb_1}+\phi_1W_{\phi_1}\bar{W}+  \pb_1W\bar{W}_{\pb_1}+\phi_1\pb_1W\bar{W}  \\
&=\frac{1}{m^2r^2_1}w_1\wbl+\frac{1}{m}(W\wbl+w_1\bar{W})+\phi_1\pb_1W\bar{W},
\end{split} \label{Vp1}
\end{equation}
and
\begin{equation}
\begin{split}
V_{\phi_2}=&W_{\phi_2}{\bar{W}}_{\pb_2}+\phi_2W_{\phi_2}\bar{W}+  \pb_2W\bar{W}_{\pb_2}+\phi_2\pb_2W\bar{W}  \\
=&\frac{1}{n^2r^2_2}w_2\wbll+\frac{1}{n}(W\wbll+w_2\bar{W})+\phi_2\pb_2W\bar{W}.
\end{split}\label{Vp2}
\end{equation}
Terms like $\phi_i\pb_iW\bar{W}$ will be dropped in the following discussions as they are several orders smaller than others.

F-term potential is separated into two parts: these independent of axions $V_1$ and these depending on axions $V_2$.
For the axion-independent part, it is
\begin{equation}
\begin{split}
V_1=&e^K\{\frac{1}{m^2r^2_1}w_1\wbl+\frac{2}{m}w_1\wbl+\frac{1}{n^2r^2_2}w_2\wbll+\frac{2}{n}w_2\wbll \\
&~~~~+2r^2(w_1\wbl+w_2\wbll)+4r(w_1\wbl+w_2\wbll)-(w_0^2+w_1\wbl+w_2\wbll)+4c^2_0\lmll\} \\
=&e^K\{(\frac{1}{m^2r^2_1}+\frac{2}{m}+2r^2+4r-1)w_1\wbl \\
&~~~~+(\frac{1}{n^2r^2_2}+\frac{2}{n}+2r^2+4r-1)w_2\wbll-w_0^2+4c^2_0\lmll+8c_0\lmll w_0\},
\end{split}
\end{equation}
while the axion-dependent part is
\begin{equation}
\begin{split}
V_2=&e^K\{(\frac{4c_0\lmll}{w_0}+\frac{1}{m}+2r-1)w_0(w_1+\wbl)+(\frac{4c_0\lmll}{w_0}+\frac{1}{n}+2r-1)w_0(w_2+\wbll) \\
&~~~~+(\frac{1}{m}+\frac{1}{n}+2r^2+4r-1)(w_1\wbll+w_2\wbl)\}.
\end{split}
\end{equation}
Terms proportional to $4c_0\lambda_2$ in $V_1$ and $V_2$ are obtained from the matter field stabilization.

To fit with observations, the parameters are taken as $m\sim O(10)$, $c_0\sim O(10^{-2})$, $\lmll\sim10^{-2}$,
$w_0\sim10^{-3}$, and $r\sim O(10)$. Hence, we have $1/m\sim4c_0\lmll/w_0\ll r$. Besides, according to Eq. (\ref{ms}),
$r=2aT_{R1}\simeq 1/2mr^2_1$, the factor $1/m^2r^2_1\simeq 2r/m\ll r^2$ is insignificant. The lower order terms will be dropped in the preliminary estimation.

Employing the formula of $w_1$ and $w_2$ in the above equations and ignoring the lower order terms,
we  rewrite scalar potentials as follows
\begin{equation}
\begin{split}
&V_1=\frac{1}{4T_{R1}T_{R2}}\{(2r^2+4r-1)A^2r_1^{\frac{2}{m}}e^{-2r} \\
&~~~~~~+(2r^2+4r-1)B^2r_2^{\frac{2}{n}}e^{-2r}-w^2_0+4c^2_0\lmll\},
\end{split}\label{Vna}
\end{equation}
and
\begin{equation}
\begin{split}
&V_2=\frac{1}{2T_{R1}T_{R2}}\{(2r-1)w_0Ar_1^{\frac{1}{m}}e^{-r}{\rm cos}(a\theta_1+b\theta_2) \\
&~~~~~~+(2r-1)w_0Br_2^{\frac{1}{n}}e^{-r}{\rm cos}(a\theta_1+b\theta_2+c\theta_2) \\
&~~~~~~+(2r^2+4r-1)ABr_1^{\frac{1}{m}}r_2^{\frac{1}{n}}e^{-2r}{\rm cos}(c\theta_2)\},
\end{split}
\end{equation}
where we have used $e^{\langle z_n\bar{z}_n\rangle}\simeq1$.

$A$ and $B$ are parameters that depend on the details of non-perturbative effects. Here we may simply assume they are close to each other, and $Ar_1^{\frac{1}{m}}e^{-r}\simeq Br_2^{\frac{1}{n}}e^{-r}\sim 10^{-4}$.

At the global minimum, the scalar potential $V_1+V_2$ decreases to
\begin{equation}
\langle V_1+V_2\rangle=\frac{1}{4\langle T_{R1}T_{R2}\rangle} (4c^2_0\lmll-w_0^2-4(2r-1)w_0Ar_1^{\frac{1}{m}}e^{-r}).
\end{equation}
Without uplifting term from matter field stabilization, the non-perturbative superpotential with no-scale-type K\"ahler potential admits an AdS vacuum, as expected.
The constant term $4c^2_0\lambda_2$ elevates the AdS vacuum to Minkowski vacuum under a constraint
\begin{equation}
4c^2_0\lambda_2=w_0^2+4(2r-1)w_0Ar_1^{\frac{1}{m}}e^{-r},
\end{equation}
and the scalar potential can be simplified as
\begin{equation}
\begin{split}
&V=2\Lambda^4_1+\Lambda^4_2+\Lambda^4_1{\rm cos}(a\theta_1+b\theta_2) \\
&~~~~~~+\Lambda^4_1{\rm cos}(a\theta_1+b\theta_2+c\theta_2)+\Lambda^4_2{\rm cos}(c\theta_2).
\end{split}
\end{equation}
Giving $w_0=2\times10^{-3}$, $T_{Ri}=20$, $r=10$, $Ar_1^{\frac{1}{m}}e^{-r}\simeq2\times10^{-4}$, we have $\Lambda^4_{1,2}\approx 10^{-8}$ which agree with the BEICEP2 observations.

Because the kinetic terms of K\"ahler moduli are non-canonical,
 the field transformations are needed to determine the physical axion decay constants. The kinetic terms are
\begin{equation}
L_K=\frac{1}{(T_i+\bar{T}_i)^2}\partial_{\mu}T_i\partial^{\mu}T_i=\frac{1}{4T^2_{Ri}}(\partial_{\mu}T_{Ri}\partial^{\mu}T_{Ri}
+\partial_{\mu}\theta_i\partial^{\mu}\theta_i).
\end{equation}
Taking field re-scale $\theta_i\rightarrow \sqrt{2}T_{Ri}\theta_i$, and using $aT_{R1}=bT_{R2}=r/2$, we get
\begin{equation}
\begin{split}
&V=2\Lambda^4_1+\Lambda^4_2+\Lambda^4_1{\rm cos}(\frac{r}{\sqrt{2}}(\theta_1+\theta_2)) \\
&~~~~~~+\Lambda^4_1{\rm cos}(\frac{r}{\sqrt{2}}(\theta_1+\theta_2+\frac{c}{b}\theta_2))+\Lambda^4_2{\rm cos}(\frac{c}{\sqrt{2}b}r\theta_2),
\end{split}
\end{equation}
where the axions $\theta_i$ now have canonical kinetic terms. Redefining the axions $\varphi_{1,2}=(\theta_1\pm\theta_2)/\sqrt{2}$, we get
\begin{equation}
\begin{split}
&V=2\Lambda^4_1+\Lambda^4_2+\Lambda^4_1{\rm cos}(r\varphi_1) \\
&~~~~~~+\Lambda^4_1{\rm cos}((1+\frac{c}{2b})r\varphi_1-\frac{c}{2b}r\varphi_2)+\Lambda^4_2{\rm cos}(\frac{c}{2b}r(\varphi_1-\varphi_2)).
\end{split}
\end{equation}
The effective mass of axion $\varphi_1$ is
\begin{equation}
m_{\varphi_1}=r\Lambda^2_1\approx 10^{-3}\gg H,
\end{equation}
where $H$ is the Hubble constant during inflation, and its value is about $10^{-4}$ in Planck units based on the BICEP2 results.
Therefore, axion $\varphi_1$ is frozen out during inflation, and another axion $\varphi_2$ drives the observed inflationary process if its decay constant
$f=2b/cr$ is of order $O(10)$.
Although $r\sim O(10)$, we have $c\ll b$, so we can easily get a large effective decay constant $f\sim 10$ by adopting a small $S_2$
symmetry breaking factor $c/b\sim10^{-2}$.

\section{conclusions}

We have proposed a concrete model to realize aligned axion inflation \cite{Kim:2004rp} for natural inflation with moduli stabilization based on two anomalous $U(1)$ gauge symmetries. String theory provides abundant axion landscapes, and the natural inflation driven by aligned axions are expected to be true of certain choices in the axion landscapes \cite{Choi:2014rja, Higaki:2014pja, Kappl:2014lra, Long:2014dta, Gao:2014uha}. For the axions as imaginary components of K\"ahler moduli, generally they appear in the potential through non-perturbative effects. Inflation driven by these axions needs subtle moduli stabilization since it requires to fix real components of moduli while keep axions sufficient light.

Similar to Refs.~\cite{Li:2014owa, Li:2014xna}, we employed the anomalous $U(1)$ gauge symmetries for moduli stabilization. K\"ahler moduli transform non-trivially under gauge symmetries $U(1)_X\times U(1)_A$, and lead to moduli-dependent FI terms in D-term potential associated with $U(1)_X\times U(1)_A$. The condensation hidden sectors $SU(n)\times SU(m)$ are assumed to have massive fundamental representations, from which the gaugino condensations introduce race-track type superpotential and cancellable D-terms. Since the D-terms depend on real components of K\"ahler moduli only, their cancellations at vacuum state lead to strong stabilizations on the real components of K\"ahler moduli.
The axions, which are imaginary components of K\'ahler moduli, remain light.

We introduced renormalizable matter couplings for matter field stabilization. Prior to D-term moduli stabilization, the matter fields have to be stabilized and obtain non-zero VEVs. This is done by Higgs-like matter couplings. Gauge symmetries $U(1)_X\times U(1)_A$ are spontaneously broken by VEVs of charged matter fields, besides, the continuous shift symmetries of K\"ahler moduli are spontaneously broken into discrete shift symmetries. Field stabilization in this model also provides a natural mechanism for uplifting the AdS vaccum to Minkowski or dS vaccum, {\it i.e.}, it introduces large positive vaccum energy, which is suitable for elevating the AdS vaccum arising from non-perturbative superpotential.

We showed that the alignment of axions in the KNP mechanism corresponds to an approximate $S_2$ symmetry between two K\"ahler moduli. The $S_2$ symmetry is approximate as it is explicitly broken by a small factor $c$. Different from the $U(1)$ sectors, the discrete $S_2$ symmetry is sustained after spontaneously gauge symmetry breaking and field stabilization. After field stabilization, the potential is determined by two axions through non-perturbative effects. The decay constants of the two axions, which are determined by the moduli stabilization and canonical field transformation, are close to $f_i=1/r$, just about the string scale.
In consequence of $S_2$ symmetry, the potential forms a steep direction along $\theta_1+\theta_2$, while its orthogonal direction $\theta_1-\theta_2$ is flat, and is suitable for inflation by taking small $S_2$ symmetry broken factor $c$.

\begin{acknowledgments}

The work of DVN was supported in part
by the DOE grant DE-FG03-95-ER-40917. The work of TL is supported in part by
    by the Natural Science
Foundation of China under grant numbers 10821504, 11075194, 11135003, and 11275246, and by the National
Basic Research Program of China (973 Program) under grant number 2010CB833000.

\end{acknowledgments}

\end{document}